\documentclass[twocolumn,showpacs,superscriptaddress]{revtex4}

\usepackage{amsmath,amssymb}
\usepackage{graphicx}
\usepackage{dcolumn}
\usepackage{bm}
\usepackage{epstopdf}
\usepackage{amstext}
\usepackage[normalem]{ulem}
\begin{document}

\title{Coherent-state-induced transparency}

\author{A. Gogyan}
\email{agogyan@gmail.com}
\author{Yu. Malakyan}
\affiliation{Institute for Physical Research, Armenian National
Academy of Sciences, Ashtarak-2, 0203, Armenia}

\date{\today}

\begin{abstract}
 We examine electromagnetically induced transparency (EIT) in an ensemble of cold $\Lambda-$type atoms induced by a quantum control field in multi-mode coherent states and compare it with the transparency created by the classical light of the same intensity. We show that the perfect coincidence is achieved only in the case of a single-mode coherent state, whereas the transparency sharply decreases, when the number of the modes exceeds the mean number of control photons in the medium. The origin of the effect is the modification of photon statistics in the control field with increasing the number of the modes that weakens its interaction with atoms resulting in a strong probe absorption. For the same reason, the probe pulse transforms from EIT-based slow-light into superluminal propagation caused by the absorption.
\end{abstract}

 \maketitle
%

\section{\protect\normalsize INTRODUCTION}

Electromagnetically induced transparency (EIT) offers unique possibilities to control optical properties  of medium with light \cite{1}. Since its first experimental observation in strontium vapors \cite{2}, it has been thoroughly investigated in different material systems including quantum dots \cite{3}, quantum wells \cite{4}, nanoplasmonics \cite{5},  metamaterials \cite{6}, and optomechanical systems \cite{7}. Many applications of EIT, ranging from reversible mapping of light-matter states in quantum memory \cite{8,9,10} to all-optical switching of one beam by another \cite{11,bajcsy,12} and cross-coupling nonlinearities at the few-photon level \cite{13,14}, reveal its ability as a basic tool for implementation of quantum information processing.  However, to date, only classical fields are employed in usual EIT schemes to yield
transparency in probe light absorption, while, despite the fundamental importance, the transparency induced in atomic ensembles by quantum fields has not been explored yet, except for the cavity-based EIT, where the classical control beam is replaced by a cavity vacuum field \cite{15}. The remarkable feature of quantum EIT-systems is their capability to generate quantum nonlinearities at the low-light level. Here we analyze the EIT created in an ensemble of $\Lambda-$type atoms by most "classical" quantum states that are multi-mode coherent states, and reveal the deviation of quantum dynamics from the classical one. We show that for the same intensity of the control field the transparency observed in the case of a single-mode coherent state perfectly coincides with the classical-field induced transparency, while the transparency substantially decreases as the number of the modes increases. As a result, the probe field changes from slow light based on the EIT to superluminal propagation caused by the absorption. This behavior results from the modification of the photon statistics in the coherent state, where the increase of mode number shifts the maximum of photon distribution to the Fock states with small number of photons provided that the total intensity of the control field remains unchanged. To enhance the atom-photon interaction and cancel the Doppler broadening, we consider propagation of a weak probe pulse in a cold atomic ensemble, such as the atoms inside a hollow-core photonic crystal fiber (HC-PCF) of a few microns in diameter, which can tightly confine both atoms and photons transversely over long interaction lengths \cite{11,bajcsy,16,17,18,19,20}. Of course, different dissipative effects are significant for coherent states with different mode numbers, however, we ignore these complications to concentrate on more fundamental issues.

In the next section, we formulate the quantum theory of EIT in $\Lambda-$type atoms and give a formal analytical solution for the probe field operators. In Sec.III, we apply this solution to the case of quantum control field in multi-mode coherent-states. Here we demonstrate how the medium transparency and dispersion are drastically altered by changing the number of the modes. We also show that the probe field group velocity changes from subluminal to superluminal propagation as the control-field mode number increases. Our conclusions are summarized in Sec. IV.

\section{FORMULATION}

We consider the transmission of the quantum probe field $\hat{E_1}$ along the z-axis in the medium of cold $\Lambda-$atoms, which resonantly interact with the probe field at the transition $1\rightarrow 3$ (Fig. 1). The atoms are driven by a quantum control field $\hat{E_2}$ tuned to the resonance at the atomic transition $2\rightarrow 3$.
We treat the problem in the one-dimensional (1D) approximation and describe the fields of carrier frequencies $\omega_i$ and wave-vectors $k_i$ by slowly varying dimensionless operators $\hat{\mathcal E}_{1,2}(z,t)$
\begin{equation}
\hat E_{i}(z,t) = \sqrt{\hbar \omega_i \over 2 \epsilon_0 V_i}\hat{\mathcal E}_{i}(z,t) \exp{[i(k_i z-\omega_i t)]} + H.c., i=1,2,
\end{equation}
where $V_i = A_i L$ is the quantization volume with $A_i$ the cross-sectional area of the $i$-th  quantized field and $L$ is the quantization length, which for simplicity is chosen below to be equal to the interaction length. The traveling-wave electric fields $\hat{\mathcal E}_{i}(z,t)$ can be expressed through single-mode operators as $\hat{\mathcal E}_{i}(z,t) = \sum_q a_i^q(t) e^{iqz} (i=1,2)$, where $a_i^q$ is the annihilation operator for the field mode with wave vector $k_i+q$. The space of the modes spanned by $q\in \{-\delta q_i/2,\delta q_i/2\}$ is bounded by $\delta q_i \leq {\Delta \omega_i \over c}$, where $\Delta \omega_i$ is the maximal bandwidth of the system at the frequency $\omega_i$. Specifically, for the probe field $\Delta \omega_1$ is the EIT window, while for the control field, for example for a transform-limited pulse, it can be chosen as $\delta q_2 = {\Delta \omega_2 \over c} = {2\pi \over c T_2}$, where $T_2$ is the control pulse duration. The single-mode operators possess the standard bosonic commutation relations $[a_i^q, a_j^{q'}] = \delta_{ij} \delta_{q,q'}$ resulting in equal space-time commutation relations
\begin{equation}
[\hat{\mathcal E}_{i}(z,t),\hat{\mathcal E}_{j}^\dagger (z,t)] = \delta_{ij} N_i,\label{comm}
\end{equation}
where $N_i$ is the total number of longitudinal modes in $i$-th field.

In the rotating wave approximation the interaction Hamiltonian is given by
\begin{equation}
H = \hbar \delta_1 \hat \sigma_{33}+\hbar \delta \hat \sigma_{22} - \hbar (g_1 \hat{\mathcal E}_{1} \hat \sigma_{31}+g_2 \hat{\mathcal E}_{2} \hat \sigma_{32} + H.c.),
\end{equation}
with $\delta = \delta_1 - \delta_2$, where $\delta_i = \omega_{3i}-\omega_i, (i=1,2)$ is the one-photon detuning of the probe and control fields, $\sigma_{\alpha \beta} = |\alpha \rangle \langle \beta |$ the atomic operators and $g_i = \mu_{3i}\sqrt{\omega_i \over 2\hbar \epsilon_0 A_iL}$ the atom-photon coupling constants, $\mu_{\alpha \beta}$ is the dipole matrix element of the atomic transition $|\alpha \rangle \rightarrow |\beta \rangle$.

The atoms are initially prepared in the state $|1 \rangle$. We solve the system equations in the weak probe-field limit. To this end, we assume that $g_1^2 I_1 \ll g_2^2 I_2$, which for $g_1\simeq g_2$ is fulfilled if $I_1 \ll I_2$. Here $I_i$ is the mean value of dimensionless intensity (photon number) operator $\hat I_i = \hat{\mathcal E}_{i}^\dagger \hat{\mathcal E}_{i}$ for the $i$-th field. Besides, the number of photons in the probe pulse is much less than the number of atoms, therefore $\sigma_{11} = 1, \sigma_{22} = \sigma_{33} \simeq 0$. Then the equations for the atomic coherences read
\begin{gather}
{\partial \over \partial t} \hat \sigma_{12} = (i\delta - \gamma_0) \hat \sigma_{12} - ig_1 \hat{\mathcal E}_{1} \hat \sigma_{32} + ig_2 \hat{\mathcal E}_{2}^\dagger \hat \sigma_{13} + \hat F_{12}, \nonumber \\
{\partial \over \partial t} \hat \sigma_{13}= -(i\delta_1 + \Gamma) \hat \sigma_{13} + ig_1 \hat{\mathcal E}_{1} + ig_2 \hat{\mathcal E}_{2} \hat \sigma_{12} + \hat F_{13}, \label{sigmas}\\
{\partial \over \partial t} \hat \sigma_{23}= -(i\delta_2 + \Gamma) \hat \sigma_{23} + ig_1 \hat{\mathcal E}_{1} \hat \sigma_{21}  + \hat F_{23}, \nonumber
\end{gather}
where $\gamma=2\Gamma$ and $\gamma_0$ are the decay rates of the excited state $|3 \rangle$ and ground-state coherence, respectively. $\hat F_{\alpha \beta}$ are $\delta$-correlated noise operators associated with relaxation.

\begin{figure}[t]
    \centering
    \includegraphics[width=0.4\textwidth]{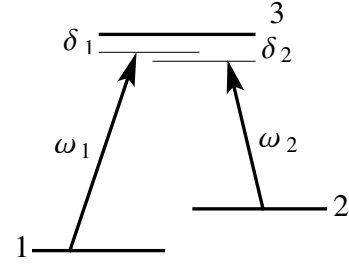}
\caption{Schematic configuration of $\Lambda-$type atoms. The probe field of frequency $\omega_1$ interacts with atoms at the $1\rightarrow 3$ transition with the detuning $\delta_1$, the quantum control field of $\omega_2$ frequency is tuned to the atomic transition $2\rightarrow 3$ with the detuning $\delta_2$.}
\end{figure}
In the slowly varying envelope approximation, the Maxwell equation for the probe-field operator has the form
\begin{equation}\label{maxwel}
\biggl( {\partial \over \partial z} + {1\over c} {\partial \over \partial t}\biggr ) \hat{\mathcal E}_{1}(z,t) = ig_1 N \hat \sigma_{13},
\end{equation}
where $N$ is the total number of atoms in the interaction region.

We consider the control field is unchanged during its propagation in the medium with light velocity $c$ that is $\hat I_2(z,t)=\hat I_2(t'),t'=t-z/c,$ owing to negligibly small excitation of the atoms to the states $|2\rangle$ and $|3\rangle$. Then we find the solution of Eqs.\eqref{sigmas} under adiabatic conditions by keeping only the first time-derivative of $\hat{\mathcal E}_{1}(z,t)$ and substitute it into Eq.\eqref{maxwel} that yields
\begin{equation}\label{maxwelsimple}
\biggl( {\partial \over \partial z} + {1\over \hat v_{g1}} {\partial \over \partial t}\biggr ) \hat{\mathcal E}_{1}(z,t) = -{g_1^2 N \tilde \gamma_0 \over c[\tilde \Gamma \tilde \gamma_0+g_2^2 \hat I_2(t')]} \hat{\mathcal E}_{1} + \hat{\mathcal F}_1(t),
\end{equation}
where $\tilde \Gamma = \Gamma+i\delta_1, \tilde \gamma_0 = \gamma_0 - i\delta$ and $\hat{\mathcal F}_1(t)$ is a commutator preserving noise operator, which, however, gives no contribution in the absence of the probe field losses \cite{21}. Since we assume that EIT conditions are fulfilled for the probe field, we neglect $\hat{\mathcal F}_1$ below. In Eq.\eqref{maxwelsimple}, the operator-valued  group velocity of the probe pulse is given by
\begin{equation}
\hat v_{g1}(z,t) = c \biggl [1+ {g_1^2 N (g_2^2 \hat I_2(t')-\gamma_0^2)  \over (\Gamma \gamma_0+g_2^2 \hat I_2(t'))^2} \biggr ]^{-1},\label{7}
\end{equation}

At this point it is convenient to take the quantum average of Eq.\eqref{maxwelsimple} with respect to the control field states, which gives
\begin{equation} \label{8}
\biggl( {\partial \over \partial z} + u {\partial \over \partial t}\biggr ) \hat{\mathcal E}_{1}(z,t) = \biggl [-{\kappa(z,t)\over 2} + i\phi(z,t) \biggr ] \hat{\mathcal E}_1(z,t),
\end{equation}
where
\begin{gather}
\kappa(z,t) = {2g_1^2 N \over c} Re[\eta(z,t)], \label{9}\\
\phi(z,t) = -{g_1^2 N \over c} Im[\eta(z,t)],\label{10}\\
\eta(z,t) = {1\over \tilde \Gamma}\langle \Psi_2| {1 \over 1+G \hat I_2(t')} |\Psi_2\rangle, \nonumber
\end{gather}
are, respectively, the linear absorption and phase-modulation coefficients of the probe field,
\begin{equation}
u(z,t) = \langle \Psi_2| {1\over \hat v_{g1}(z,t)} |\Psi_2\rangle,\label{11}
\end{equation}
is the mean value of the inverse group-velocity operator, $|\Psi_2\rangle$ is the input state of the control field and $G = g_2^2/(\tilde \Gamma \tilde \gamma_0)$. By replacing the operator $\hat I_2(t)$ in Eqs.(\ref{7}-\ref{10}) with its expectation value such that $g_2^2 \hat I_2(t) \rightarrow \Omega_c^2(t)$ where $\Omega_c(t)$ is the Rabi frequency of the control field, the classical expression for the probe absorption \cite{1} is reproduced
\begin{equation}
\kappa_{cl}(t) = {2g_1^2 N \over c} Re[\frac{\tilde \gamma_0}{\tilde \Gamma \tilde \gamma_0+\Omega_c^2(t)}], \label{12p}
\end{equation}
In the limit of $v_1\ll c$, the solution of Eq.\eqref{8} can be expressed in terms of retarded time $\tau = t-zu$ as
\begin{equation}
\begin{split}
\hat{\mathcal E}_{1}(z,t)& = \hat{\mathcal E}_{1}(0,\tau) \exp \biggl [ -\int \limits_0 \limits^z dz'\biggl ({1\over 2} \kappa (\tau+z'u)  \\
& + i\phi (\tau+z'u)\biggr )\biggr ].\label{12}
\end{split}
\end{equation}
This is the central result of this paper and is applicable for any input state of the control field. As to the probe field, we consider an input single-photon wave packet with duration $T_1>\gamma^{-1}$ to guarantee the weak field and adiabatic approximations.

In the next section we use this solution to explore the quantum dynamics of the system in the case of multi-mode coherent state of the control field.

\section{CONTROL FIELD IN MULTI-MODE COHERENT STATE}

In this section we study the transparency of the atomic sample induced by the control field, which is in the multi-mode coherent state $|\Psi_2\rangle = |\alpha \rangle = \prod_q |\alpha_2^{(q)} \rangle$, where the state $|\alpha_2^{(q)} \rangle$ is a coherent state of $q$-th mode $|\alpha_2^{(q)} \rangle = e^{-|\alpha_q|^2/2} e^{\alpha_q a_2^{q\dagger}}|0\rangle$. Therefore, the state $|\alpha \rangle$  is the eigenstate of the input operator $\hat{\mathcal E}_2(0,t)$ at $z=0$ with the eigenvalue $\alpha(t) = \sum_q \alpha_q e^{-icqt}$: $\hat{\mathcal E}_2(0,t) |\alpha\rangle = \alpha(t)|\alpha\rangle$. Upon propagating through the medium, the probe pulse experiences linear attenuation and phase modulation, which are described by the functions $\kappa(x)$ and $\phi(x)$ in Eq.\eqref{12}, respectively.

The mean values of the operators in Eqs.(\ref{9},\ref{10}) are found by expanding $\eta(\tau)$ into power series of $\hat I_2$
\begin{equation}
\begin{split}\label{13}
D(\tau) &=\langle \alpha| {1\over 1+G \hat I_2(\tau)}|\alpha \rangle \\
&= \langle \alpha| \sum \limits_m (-1)^m G^m (\hat{\mathcal E}_{2}^\dagger(\tau) \hat{\mathcal E}_{2}(\tau))^m|\alpha \rangle
\end{split}
\end{equation}
and applying the commutation relation \eqref{comm}. The straightforward calculations yield
\begin{equation}
\begin{split}\label{14}
D(\tau)& = \exp \biggl ( -{|\alpha(\tau)|^2 \over N_2} \biggr ) \sum_k {\biggl ({|\alpha(\tau)|^2 \over N_2}\biggr)^k {1\over k!} {1\over 1+G N_2 k} } \\
& =\exp \biggl ( -{|\alpha(\tau)|^2 \over N_2} \biggr ) {{_1}F{_1}}\biggl [ {1\over GN_2},1+{1\over GN_2},{|\alpha(\tau)|^2 \over N_2}\biggr ],
\end{split}
\end{equation}
where $_1F_1$ is the Kummer confluent hypergeometric function.

In Eq.\eqref{14} $|\alpha(\tau)|^2$ describes the mean number of control photons in the interaction region and $|\alpha(\tau)|^2/N_2$ is the mean photon number per mode. For slowly varying control intensity with the pulse width $T_2$ exceeding the interaction time of the probe pulse with the medium: $T_2 > T_1+uL$, the Eq.\eqref{12} takes a simple form
\begin{equation}
\hat{\mathcal E_1}(z,t) = \hat{\mathcal E_1}(0,\tau)\exp{[(-{\kappa \over 2}+i\phi)z]},
\end{equation}
with
\begin{equation}\label{16}
\kappa = {2g_1^2N\over c} Re \bigl[{D \over \tilde \Gamma}], \quad \phi =-{g_1^2N\over c} Im \bigl[{D \over \tilde \Gamma}].
\end{equation}
where $D$ is given by Eq.\eqref{14} for the constant mean photon number $|\alpha(\tau)|^2\equiv |\alpha|^2$.
\begin{figure}[t]
    \centering
        \includegraphics[width=0.4\textwidth]{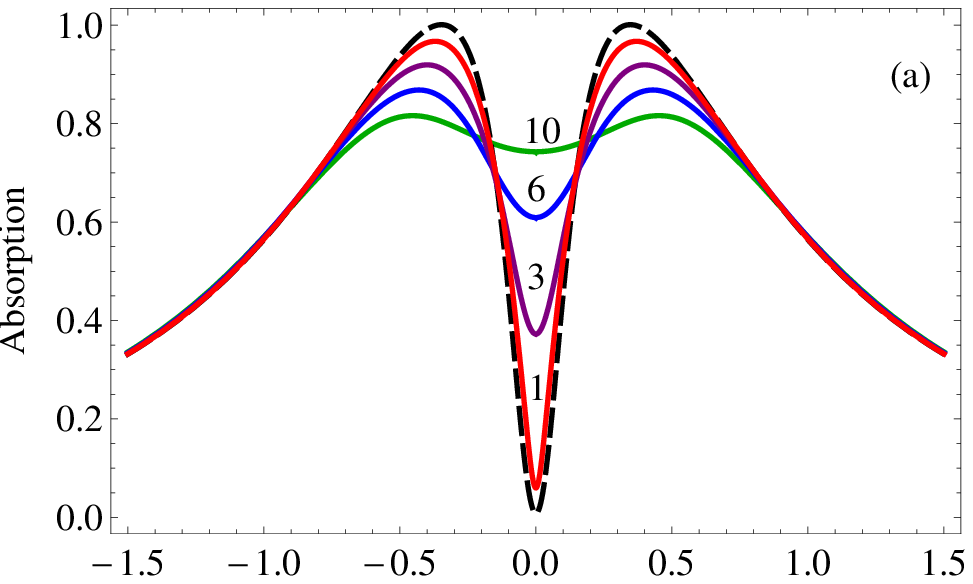}
        \includegraphics[width=0.4\textwidth]{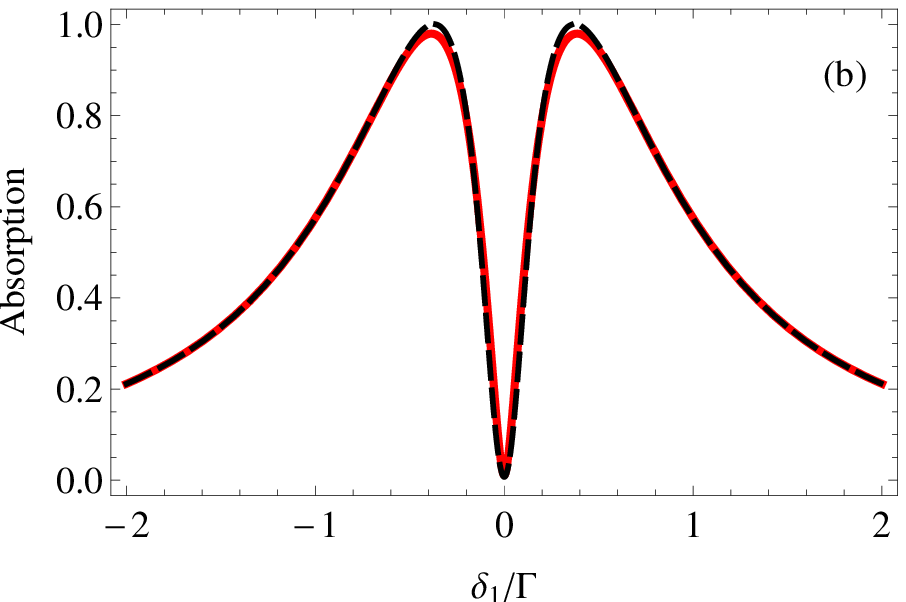}
    \caption{(Color online) (a) Probe absorption spectrum induced by the quantum control field in multi-mode coherent states for $g_2=0.5\Gamma,|\alpha|^2=3$, and $\delta_2 = 0$ and different mode numbers 1(red), 3(purple), 6(blue), and 10(green) shown on the solid curves. The dashed line is the probe absorption with classical control field of the same intensity;  (b) probe absorption for $g_2=0.06\gamma$ induced by a coherent state with $N_2=10$ and $|\alpha|^2=36$ (solid red line) and classical control field with the corresponding intensity (black dashed line).}
\end{figure}

In Fig. 2a we plot the probe absorption spectrum $\kappa(\delta_1)$ in units of $2g_1^2N/(c\Gamma)$ for different numbers of coherent-state modes and compare it with the spectrum $\kappa_{cl}(\delta_1)$ obtained from Eq.\eqref{12p} for classical control field of the same intensity $\Omega_c^2 = g_2^2 \langle \alpha| \hat I_2|\alpha\rangle = g_2^2 |\alpha|^2$. We consider the configuration of transversely trapped atoms inside the HC-PCF, which has been experimentally realized in \cite{11,bajcsy}. The atoms are confined in the radial direction with a radius smaller than the radius of the fiber core. This transverse confinement prevents atom-wall collisions inside the fiber core making the atomic ground-state decoherence due to atom-wall collisions negligibly small \cite{11}. Nevertheless, for correct numerical calculations, we use the finite value of the decoherence rate $\gamma_0 =10^{-3}$MHz. The rest parameters correspond to experimental conditions in \cite{11}: atomic excited-state decay rate $\gamma/2\pi=6$ MHz, field wavelengths $\lambda_1\simeq \lambda_2=800$ nm, cross-sectional area $A_1=A_2=3\times10^{-7} \text{cm}^2$, number of atoms $N=10^3$, interaction length $L=3$cm. We consider a weak coherent state  $|\alpha|^2=3$, which can be experimentally obtained by controlled attenuation of a small portion of the laser emission \cite{22}. It is evident from Fig. 2 that in the quantum regime the transparency of the medium reduces, as the number of coherent modes increases. This immediately follows also from Eq.\eqref{14}, where the function $_1F_1$ and the exponential factor tend to unity for $|\alpha|^2/N_2 < 1$, while for the transparency one needs $D \ll 1$ as seen from Eqs.\eqref{16}. For large $N_2$ this strongly weakens the interaction of the control field with atoms, so that the maximum transparency is expected in the case of single-mode coherent state or CW control field, when the transparency in fact perfectly coincides with that induced by the classical control field (Fig.2a, red line). To understand the origin of this behavior, let us recall that each mode $|\alpha_2^{(q)} \rangle$ of the input coherent state of the control field is itself a coherent superposition of many Fock components including the vacuum state. From the structure of the Fock-state dependence in Eq.\eqref{14} we recognize that the total contribution to the EIT coming from all the modes with a given number of photons $k$ is determined by the term $GN_2k$ in the denominator of the right hand side of the first equation of \eqref{14} indicating that the greater transparency is expected for larger $k$. However, for fixed $|\alpha|^2$ and large $N_2$, the transparency is induced mainly by the lower Fock states with small number of photons $k\sim 1$, because the probability distribution of Fock-states in $D$ is defined by the $k$-th degree of the mean photon number per mode $(|\alpha|^2 / N_2)^k$, which quickly decreases with increasing $k$. Moreover, the probe field experiences huge absorption, when the control field is in the vacuum state $k=0$. As a result, the EIT peak decreases as $N_2$ increases by the law shown in Fig.3. Thus, the observed effect is due to modification of photon statistics in the coherent state, where the increase of mode number $N_2$, while keeping the intensity $|\alpha|^2$ of the control field unchanged, shifts the maximum of photon distribution to the Fock states with smaller number of control photons $k\sim 1$. Obviously, this stringent quantum behavior emerges for small values of $|\alpha|^2$ and disappears in the classical limit $|\alpha|^2\gg N_2$. In the last case, the perfect EIT is displayed in Fig.2b for $|\alpha|^2=36$ and $N_2=10$, where atom-photon coupling $g_2$ is taken small enough to remain below the Autler-Townes regime.

\begin{figure}[h!]
    \centering
        \includegraphics[width=0.4\textwidth]{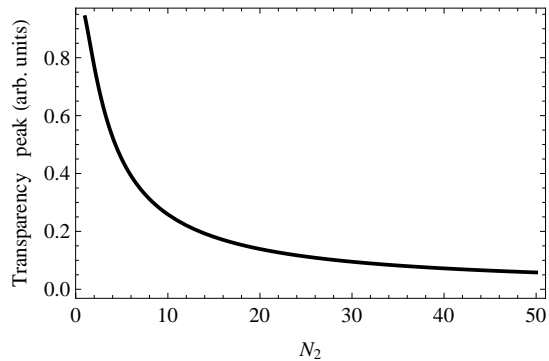}
    \caption{Transparency peak as a function of the coherent-state mode-number $N_2$ for the parameters as in the Fig.2 normalized to that of $N_2=1$.}\label{n2}
\end{figure}
\begin{figure}
    \centering
        \includegraphics[width=0.4\textwidth]{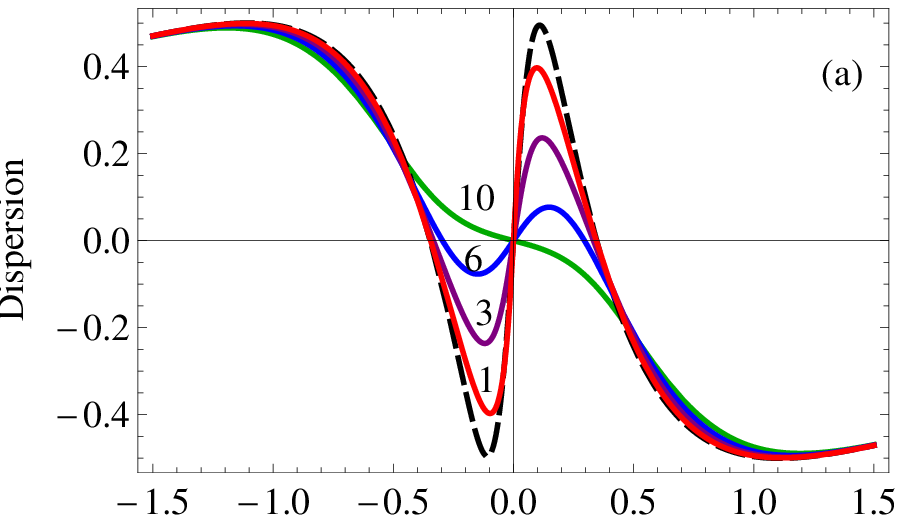}
        \includegraphics[width=0.4\textwidth]{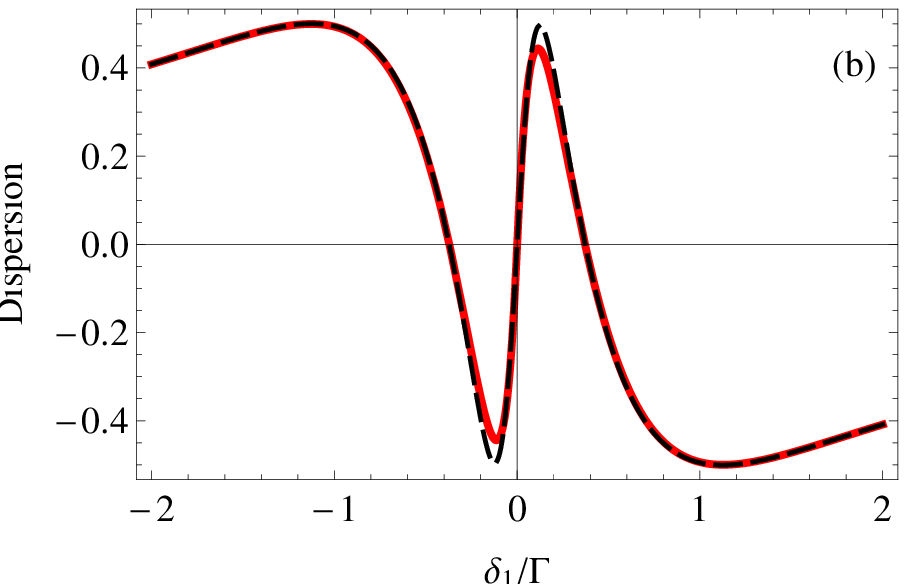}
    \caption{(Color online)(a) Probe dispersion spectra for different mode numbers of the control field for the parameters as in Fig.2; (b) dispersion spectra for the two cases shown in Fig.2b.}\label{dispersion}
\end{figure}
\begin{figure}[b]
    \centering
        \includegraphics[width=0.4\textwidth]{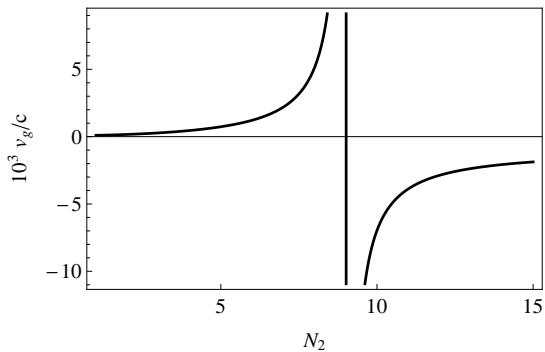}
    \caption{The switching of the probe group velocity from subluminal to superluminal propagation in vicinity of $N_2\simeq 7$. The  parameters are the same as those in Fig. 2.}\label{groupvel}
\end{figure}
Much physical insight into mechanisms that depress the effect of the control field is available from the probe field dispersion $\phi(\delta_1)$ presented in Fig.4 in units of $g_1^2N/(c\Gamma)$. It is seen that the normal dispersion is switched into anomalous dispersion (green curve) via increasing the mode number, so the probe field can change from slow light based on the EIT to superluminal propagation caused by the absorption. To understand this effect we analyzed the average group velocity of the probe pulse defined as $v_{g1} = {1\over u} = {c\over n_g}$ with the group index
\begin{equation}
n_g = \langle \Psi_2|{g_1^2 N(g_2^2 \hat I_2-\gamma_0^2)\over (\Gamma \gamma_0 +g_2^2 \hat I_2)^2}|\Psi_2\rangle. \label{groupindex1}
\end{equation}
Calculating analogously to $\kappa(x), \phi(x)$ in Eqs.\eqref{16}, we get
\begin{equation}
n_g = g_1^2 N e^{-{|\alpha|^2\over N_2}}\sum \limits_{k=0}^\infty \biggl ( {|\alpha|^2\over N_2}\biggr )^k {1\over k!}{g_2^2 N_2 k -\gamma_0^2\over (\Gamma \gamma_0 +g_2^2 N_2 k)^2},\label{groupindex}
\end{equation}
showing that when the control field is in the vacuum state $k=0$, the probe group index is negative, which, however, for the moderate values of $N_2$ is entirely compensated by the contribution of occupied Fock states with $k\geq 1$. Yet, at larger values of $N_2$ the compensation is not complete and, therefore, when the mode number increases from one, the probe group velocity first increases from $v_{g1}=10km/s$ and then upon passing the infinite value $v_{g1}\rightarrow \infty$ becomes negative. Fig.5 shows the change of probe velocity from sub- to superluminal propagation, which for our parameters occurs approximately at $N_2\simeq 7$.

Finally, we discuss the connection between the classical and quantum descriptions of the EIT represented by Eqs.\eqref{12p} and \eqref{16}, respectively. In the first case, the probe absorption depends on the Rabi frequency or electric-field amplitude of the control beam and, hence, on the total number of control photons. On the contrary, in the quantum regime, the perfect EIT is induced by several control photons in the interaction region. It is clear, however, that the latter occurs only if the few-photon level in the interaction region is supported for a long time until the probe pulse propagates through the medium. In the traveling-wave geometry, this can be achieved by a constant flow of control photons from outside. For flying coherent-state, the photon flux into interaction volume is readily calculated to be $f(\tau) =\frac{c}{L}|\alpha(\tau)|^2$, from which the total number of control photons is obtained as $n_c=\int f(t)dt$. For our parameters including $T_2\sim 1\mu s>T_1$ one finds $n_c\sim 3\times 10^4$, which is comparable with that of a classical control field used in the experiment in Ref.\cite{11}. Thus, in the considered configuration, to induce the same transparency in both classical and quantum cases almost the same amount of control photons is required. However, it is possible to implement the quantum regime of the EIT on the basis of the presented theory without using the multiphoton control pulse, if the medium is placed inside an optical ring cavity and a quantum control field containing just a few photons is used as the circulating intracavity field provided that the cavity decay time is larger as compared to the probe transmission time through the medium.

\section{CONCLUSIONS}

In this paper, we have shown that the EIT system is a promising platform to study the quantum properties of multi-mode coherent states, which to our knowledge, have never been explored in previous studies.  Our main conclusion is that the quantum nature of these states is manifested, when the number of the modes exceeds the mean number of control photons in the medium. Otherwise, the classical result is reproduced. In the cold-atom based EIT scheme considered here, this conclusion is confirmed by the peculiar behavior of the absorption and group velocity of the probe pulse. An important result of our study is the general solution for the probe pulse propagation in the quantum regime of the EIT obtained for arbitrary input quantum state of the control field. The use of interaction between the single-photon probe pulse and quantum control fields opens possibilities to study the spatiotemporal structure and quantum statistics of propagating multiphoton states, which is a challenging task currently. In our scheme the control field not only induces transparency for the quantum probe pulse, but simultaneously imparts on the latter a phase shift at the level of few photons via the Kerr-type nonlinear interaction between the two fields. This is fundamentally different from most studies of cross-phase shifts of quantum light, which are based on $N$-type EIT medium \cite{14,23,24,26,27}, where the Kerr interaction between the signal and probe fields is mediated by a strong classical field.
The measurement of the phase shift induced by the control field on the single-photon probe pulse provides nondestructive detection of the number of control photons, thus enabling the quantum nondemolition measurement of the photon number \cite{28} in the control beam. For these measurements, the scheme of optical ring cavity is more preferable. The conditional phase shifts can be also used as quantum phase gates for quantum logic operations \cite{29} given that there are large susceptibilities in the control channel to retain the weak-field approximation. Future work will explore these issues for new multiphoton states (see, for instance, \cite{30}) of the control field, which are needed for modern quantum information experiments.

\subsection*{Acknowledgments}

This research has been conducted in the scope of the International Associated Laboratory (CNRS-France and
SCS Armenia) IRMAS. We acknowledge additional support from the European Union Seventh Framework Programme
Grant No.609534-SECURE-R2I.


\end{document}